\documentclass[twoside]{ilcws07}
\usepackage[latin1]{inputenc}
\usepackage[dvips]{graphicx,epsfig,color}
\usepackage{wrapfig,rotating}
\usepackage{amssymb,amsmath,array}

\def\beq{\begin{equation}}
\def\eeq{\end{equation}}
\def\beqar{\begin{eqnarray}}
\def\eeqar{\end{eqnarray}}
\def\barr#1{\begin{array}{#1}}
\def\earr{\end{array}}
\def\bfi{\begin{figure}}
\def\efi{\end{figure}}
\def\btab{\begin{table}}
\def\etab{\end{table}}
\def\bce{\begin{center}}
\def\ece{\end{center}}

\def\text{\textstyle}

\def\ga{\gamma}

\def\mathswitchr#1{\relax\ifmmode{\mathrm{#1}}\else$\mathrm{#1}$\fi}

\newcommand{\PW}{\mathswitchr W}
\newcommand{\PZ}{\mathswitchr Z}

\newcommand{\PH}{\mathswitchr H}

\newcommand{\Pep}{\mathswitchr {e^+}}
\newcommand{\Pem}{\mathswitchr {e^-}}

\def\mathswitch#1{\relax\ifmmode#1\else$#1$\fi}

\begin{document}
\title{\boldmath{
Precision calculations for $\PH\to\PW\PW/\PZ\PZ\to4$fermions 
with {\sc PROPHECY4f}
}} 
\author{A.~Bredenstein$^1$, A.~Denner$^2$, S.~Dittmaier$^3$
and M.M.~Weber$^4$
\thanks{Supported in part by the European
Community's Marie-Curie Research Training Network HEPTOOLS under
contract MRTN-CT-2006-035505.
}
\vspace{.3cm}\\
1- High Energy Accelerator Research Organization (KEK) \\
Tsukuba, Ibaraki 305-0801, Japan
\vspace{.1cm}\\
2- Paul Scherrer Institut \\
W\"urenlingen und Villigen, CH-5232 Villigen PSI, Switzerland
\vspace{.1cm}\\
3- Max-Planck-Institut f\"ur Physik (Werner-Heisenberg-Institut) \\
D-80805 M\"unchen, Germany
\vspace{.1cm}\\
4- Department of Physics, University at Buffalo \\
The State University of New York, Buffalo, NY 14260-1500, USA
}

\maketitle

\begin{abstract}
{\sl PROPHECY4f\/} is a Monte Carlo event generator for precise simulations 
of the Higgs-boson decay $\PH\to\PZ\PZ/\PW\PW\to4$fermions, 
supporting leptonic, semileptonic, and four-quark final states.
Both electroweak and QCD corrections are included.
Treating the intermediate gauge bosons as resonances,
the calculation covers the full Higgs-boson mass
range above, near, and below the gauge-boson pair thresholds. 
In this article we pay particular attention to the recently
implemented option of {\sl PROPHECY4f\/} to generate unweighted events.
\end{abstract}

\section{Introduction}

The decay of a Standard Model Higgs boson into weak-boson pairs
with a subsequent decay into four fermions,
$\PH\to\PZ\PZ/\PW\PW\to4f$, plays an important role
both in the Higgs search at the LHC \cite{Asai:2004ws} and in
precision Higgs physics at the planned 
International $\rm{e}^+\rm{e}^-$ Linear 
Collider (ILC). The spin and the CP
properties of the Higgs boson could be verified upon studying
angular and invariant-mass distributions \cite{Barger:1993wt}
of the decay fermions. In order to match the estimated experimental
precision in predictions, a Monte Carlo generator for
$\PH\to\PZ\PZ/\PW\PW\to4f$ including radiative corrections is
needed.
In the past, only the electroweak ${\cal O}(\alpha)$ corrections to
decays into on-shell gauge bosons $\PH\to\PZ\PZ/\PW\PW$
\cite{Fleischer:1980ub} and some leading higher-order corrections were
known. However, in the threshold region the on-shell approximation
becomes unreliable.  Below the gauge-boson-pair thresholds only the
leading order was known until recently.

{\sl PROPHECY4f} \cite{Bredenstein:2006rh} is a recently constructed
Monte Carlo event generator for $\PH\to\PZ\PZ/\PW\PW\to4f$ that
includes electroweak and QCD corrections as well as some higher-order
improvements. Since the process with off-shell gauge bosons is
consistently considered without any on-shell approximations, the
obtained results are valid above, near, and below the gauge-boson pair
thresholds. 
In this note we briefly describe the structure 
of the underlying calculations and illustrate
the new option of {\sl PROPHECY4f\/} to generate 
unweighted events by reproducing some of the numerical results
presented in Ref.~\cite{Bredenstein:2006rh}.

\section{Calculational details}

The calculation of the complete electroweak ${\cal O}(\alpha)$ 
and strong ${\cal O}(\alpha_{\mathrm{s}})$
corrections to the processes $\PH\to4f$, which includes both the 
corrections to the decays $\PH\to\PZ\PZ\to4f$ and
$\PH\to\PW\PW\to4f$ and their interference,
is described in Ref.~\cite{Bredenstein:2006rh} in detail.
Each ingredient of the calculation has been worked out twice, using
independent approaches as far as possible.

For the implementation of the finite widths of the gauge bosons we use
the ``complex-mass scheme'', which was introduced in
Ref.~\cite{Denner:1999gp} for lowest-order calculations and
generalized to the one-loop level in Ref.~\cite{Denner:2005fg}.  In
this approach the W- and Z-boson masses are consistently considered as
complex quantities, defined as the locations of the propagator poles
in the complex plane.  The scheme fully respects all relations that
follow from gauge invariance.  

The one-loop amplitudes of the virtual corrections
have been generated with {\sl FeynArts}, using the two
independent versions 1 ~\cite{Kublbeck:1990xc} and 3~\cite{Hahn:2000kx}.
They have been generated and evaluated both in the
conventional 't~Hooft--Feynman gauge and in the background-field
formalism using the conventions of Refs.~\cite{Denner:1991kt} and
\cite{Denner:1994xt}, respectively. 
One version of the algebraic part of the calculation
is based on an in-house program
implemented in {\sl Mathematica}, another has been completed with
the help of {\sl FormCalc} \cite{Hahn:1998yk}.  
The one-loop tensor integrals are evaluated as in the calculation of the
corrections to ${\rm e}^+{\rm e}^-\to4\,$fermions
\cite{Denner:2005fg,Denner:2005es}.  They are recursively reduced to
master integrals at the numerical level.  The scalar master integrals
are evaluated for complex masses using the methods and results of
Refs.~\cite{'tHooft:1979xw}. 
Tensor and scalar 5-point functions are
directly expressed in terms of 4-point integrals \cite{Denner:2002ii}.
Tensor 4-point and 3-point integrals are reduced to scalar integrals
with the Passarino--Veltman algorithm \cite{Passarino:1979jh} as long
as no small Gram determinant appears in the reduction. If small Gram
determinants occur, the alternative reduction schemes of 
Ref.~\cite{Denner:2005nn} are applied.

Since corrections due to the self-interaction of the Higgs boson
become important for large Higgs masses, we have included the
dominant two-loop corrections to the decay ${\rm H}\to VV$
proportional to $G_\mu^2 M_{\rm H}^4$ in the large-Higgs-mass limit
which were calculated in Ref.~\cite{Ghinculov:1995bz}.

The soft and collinear singularities appearing in the real corrections
are treated both in the dipole subtraction approach
\cite{Dittmaier:2000mb} and in the phase-space slicing method.
For the calculation of
non-collinear-safe observables we use the extension of the subtraction
method introduced in Ref.~\cite{Bredenstein:2005zk}.
Final-state radiation off charged leptons beyond ${\cal
O}(\alpha)$, which is relevant if bare lepton momenta enter
the event selection, is supported for weigthed events only.
These corrections~\cite{Bredenstein:2006rh} are sizeable only in regions
where the lowest-order prediction is relatively small and
can amount to 4\% for muons and up to about 10\% for electrons.

\section{Event generation}

{\sl PROPHECY4f\/} employs a multi-channel Monte Carlo generator 
similar to {\sl RacoonWW} \cite{Denner:1999gp,Denner:2002cg} and {\sl
Coffer}$\ga\ga$ \cite{Bredenstein:2005zk,Bredenstein:2004ef}.
The results obtained this way have been checked using
the adaptive integration program {\sl VEGAS}
\cite{Lepage:1977sw}.
In its default version {\sl PROPHECY4f\/} generates weighted events,
which are not positive definite. 

As a new option, the program
now supports the generation of unweighted events in its
``phase-space-slicing'' branch, applying
a hit-and-miss algorithm similar to the one used by {\sl RacoonWW}.
Each time an unweighted event is generated, a Fortran subroutine is
called where information about the event is provided in the format of the
Les Houches Accord \cite{Boos:2001cv} (Fortran common block {\tt
HEPEUP}). This subroutine can be modified by the user in order to read
out the events.

In the unweighting procedure also negative events occur. Although their
number is reduced by using only the sum of the tree-level, the virtual,
and the soft endpoint contribution, they cannot be avoided completely. 
In {\sl PROPHECY4f\/} the remaining negative events are treated 
in the same way as the positive events, i.e.\ they can be read out by the
user in a subroutine. Their contribution ranges from less than a per
mille to slightly more than one per cent of all events, depending on the
Higgs-boson mass. 

The price for generating unweighted events is an increase of CPU time
by about a factor $10^2$ up to some $10^3$ w.r.t.\ weighted-event
generation, depending on the chosen $4f$ final state and the Higgs-boson
mass. The results compared below are obtained with $5\times10^5$
unweighted and $5\times10^7$ weighted events. The generation of
these unweighted events requires about 2 days on a AMD Opteron 252
2.6GHz CPU. 
However, one should keep in mind that such unweighted decay events
could be generated once for a chosen setup and stored in a database.
Simulations of Higgs production at the LHC or ILC could then
just randomly pick events for the Higgs decays from the database.

\section{Numerical results}

The input parameters and the details of the setup in our numerical
evaluation are provided in Ref.~\cite{Bredenstein:2006rh}, where
a comprehensive survey of numerical results is presented. The
results shown in the following are obtained without applying photon
recombination, i.e.\ invariant masses and angles are derived from
bare lepton momenta.

\begin{figure}
\setlength{\unitlength}{1cm}
\centerline{
\begin{picture}(7.1,7.7)
\put(-1.7,-14.5){\includegraphics{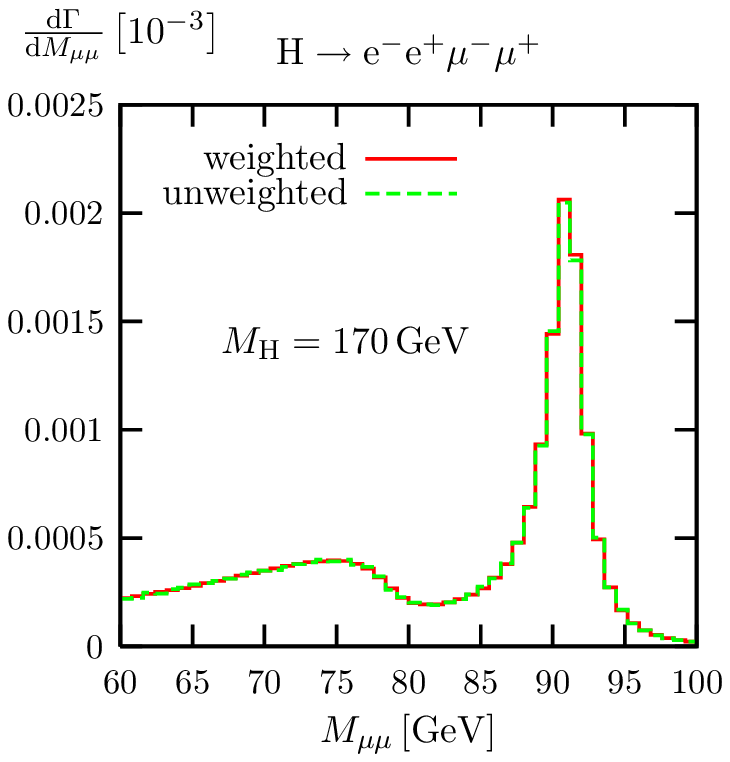}}
\end{picture}
\begin{picture}(7.0,7.7)
\put(-1.7,-14.5){\includegraphics{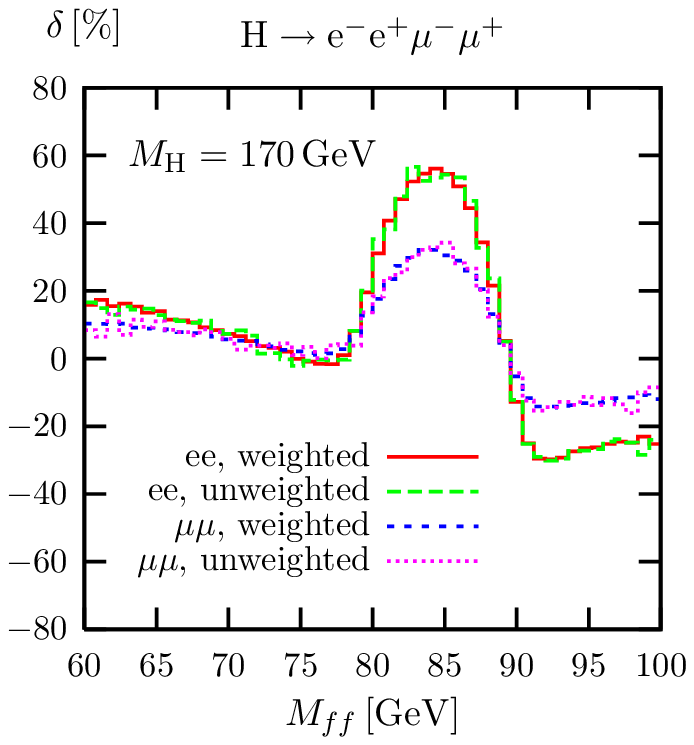}}
\end{picture} } 
\vspace*{-1em}
\caption{Corrected distribution in the invariant mass of the $\mu^-\mu^+$ pair
  (l.h.s.) and relative corrections for $\Pem\Pep$ and $\mu^-\mu^+$ pairs 
  (r.h.s.) in the decay $\PH\to\Pem\Pep\mu^-\mu^+$, obtained with
  weighted and unweighted events.}
\label{fig:inv}
\end{figure}

\begin{figure}
\setlength{\unitlength}{1cm}
\centerline{
\begin{picture}(7.1,7.7)
\put(-1.7,-14.5){\includegraphics{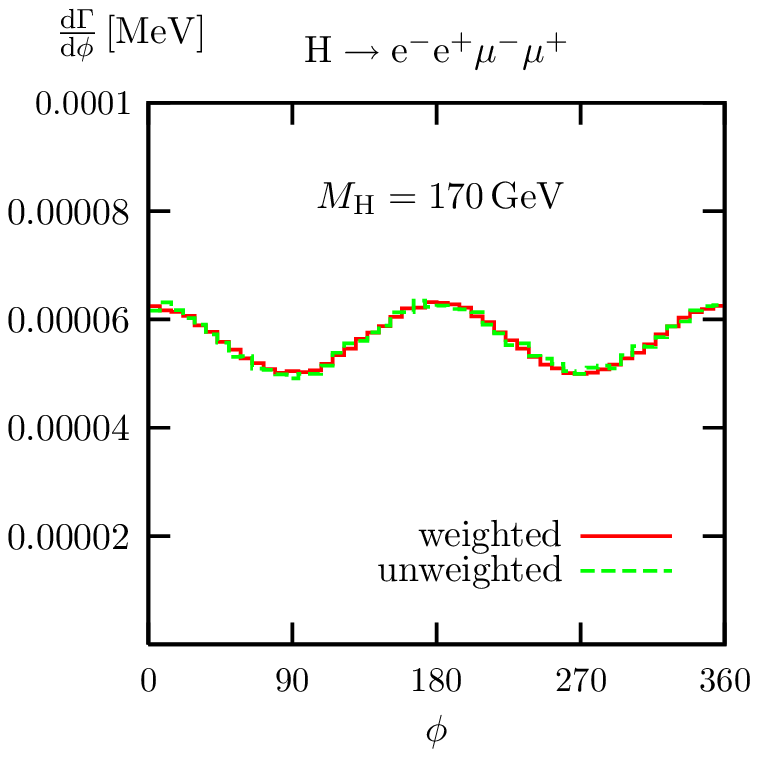}}
\end{picture}
\begin{picture}(7.0,7.7)
\put(-1.7,-14.5){\includegraphics{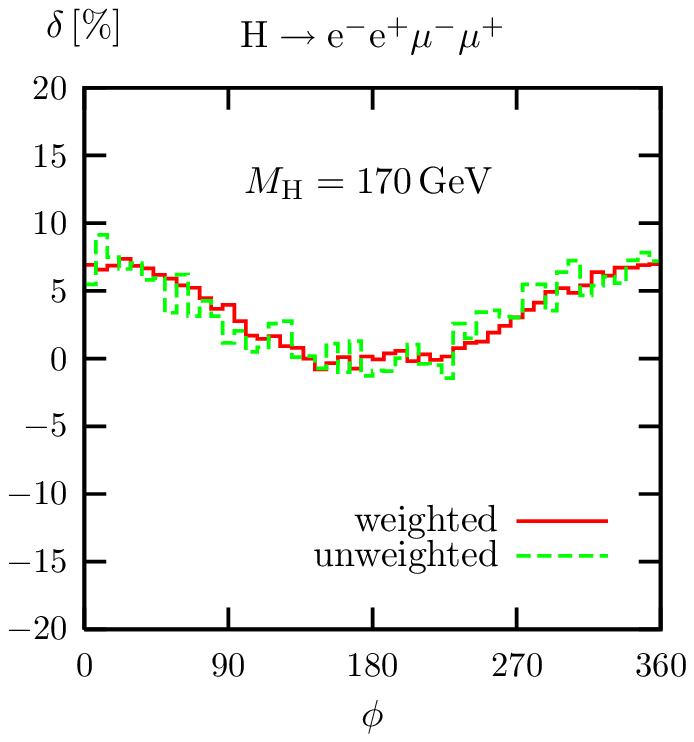}}
\end{picture} } 
\vspace*{-1em}
\caption{Corrected distribution in the angle between the $\PZ\to l^-l^+$
  decay planes in the Higgs rest frame (l.h.s.) and relative corrections 
  (r.h.s.) in the decay $\PH\to\Pem\Pep\mu^-\mu^+$, obtained with
  weighted and unweighted events.}
\label{fig:phi}
\end{figure}

In this brief article we focus only on the decay
$\PH\to\Pem\Pep\mu^-\mu^+$ and show the distributions in the
invariant masses of the decay leptons and the angle between the 
Z-decay planes in Figs.~\ref{fig:inv} and \ref{fig:phi}, respectively.
These distributions play an important role in the verification
of the discrete quantum numbers of the Higgs boson
\cite{Barger:1993wt}.
Since the radiative corrections significantly distort the distributions,
they have to be taken into account if these observables are used
to set bounds on non-standard couplings.
Neglecting the corrections could result in faking new-physics effects.
A detailed discussion of the corrections to these distributions
can be found in Ref.~\cite{Bredenstein:2006rh}.
Here we merely emphasize the agreement between the results obtained
with weighted and unweighted events generated with {\sl PROPHECY4f}.

\section{Conclusions}

The generator {\sl PROPHECY4f\/}, which simulates the
Higgs decays ${\rm H}\to\PZ\PZ/\PW\PW\to 4f$ including electroweak
and QCD corrections at the state of the art, is extended by an option
for the generation of unweighted events. 
The consistency of the new option is illustrated in 
invariant-mass and angular distributions.


\begin{thebibliography}{99}

\bibitem{Asai:2004ws}
  S.~Asai {\it et al.},
  Eur.\ Phys.\ J.\ C {\bf 32S2} (2004) 19
  [hep-ph/0402254];
%
  S.~Abdullin {\it et al.},
  Eur.\ Phys.\ J.\ C {\bf 39S2} (2005) 41.

\bibitem{Barger:1993wt}
  V.D.~Barger, K.M.~Cheung, A.~Djouadi, B.A.~Kniehl and P.M.~Zerwas,
  Phys.\ Rev.\ D {\bf 49} (1994) 79
  [hep-ph/9306270]; 
%
  S.Y.~Choi, D.J.~Miller, M.M.~M\"uhlleitner and P.M.~Zerwas,
  Phys.\ Lett.\ B {\bf 553} (2003) 61
  [hep-ph/0210077];
%
  C.~P.~Buszello, I.~Fleck, P.~Marquard and J.~J.~van der Bij,
  Eur.\ Phys.\ J.\  C {\bf 32} (2004) 209
  [arXiv:hep-ph/0212396].

\bibitem{Fleischer:1980ub}
  J.~Fleischer and F.~Jegerlehner,
  Phys.\ Rev.\ D {\bf 23} (1981) 2001;
%
  B.A.~Kniehl,
  Nucl.\ Phys.\ B {\bf 352} (1991) 1 and
%
  Nucl.\ Phys.\ B {\bf 357} (1991) 439;
%
  D.Y.~Bardin, P.K.~Khristova and B.M.~Vilensky,
  Sov.\ J.\ Nucl.\ Phys.\  {\bf 54} (1991) 833
  [Yad.\ Fiz.\  {\bf 54} (1991) 1366].

\bibitem{Bredenstein:2006rh}
  A.~Bredenstein, A.~Denner, S.~Dittmaier and M.~M.~Weber,
  Phys.\ Rev.\  D {\bf 74} (2006) 013004
  [hep-ph/0604011] and
%
  JHEP {\bf 0702} (2007) 080
  [hep-ph/0611234].

\bibitem{Denner:1999gp}
  A.~Denner, S.~Dittmaier, M.~Roth and D.~Wackeroth,
  Nucl.\ Phys.\ B {\bf 560} (1999) 33
  [hep-ph/9904472].
 
\bibitem{Denner:2005fg}
  A.~Denner, S.~Dittmaier, M.~Roth and L.~H.~Wieders,
  Nucl.\ Phys.\  B {\bf 724} (2005) 247
  [hep-ph/0505042].

\bibitem{Kublbeck:1990xc}
J.~K\"ublbeck, M.~B\"ohm and A.~Denner,
Comput.\ Phys.\ Commun.\  {\bf 60} (1990) 165;
H.~Eck and J.~K\"ublbeck, {\it Guide to FeynArts 1.0\/},
University of W\"urzburg, 1992.

\bibitem{Hahn:2000kx}
T.~Hahn,
Comput.\ Phys.\ Commun.\  {\bf 140} (2001) 418
[hep-ph/0012260].

\bibitem{Denner:1991kt}
  A.~Denner,
  Fortsch.\ Phys.\  {\bf 41} (1993) 307.

\bibitem{Denner:1994xt}
  A.~Denner, G.~Weiglein and S.~Dittmaier,
  Nucl.\ Phys.\ B {\bf 440} (1995) 95
  [hep-ph/9410338].

\bibitem{Hahn:1998yk}
T.~Hahn and M.~Perez-Victoria,
Comput.\ Phys.\ Commun.\  {\bf 118} (1999) 153
[hep-ph/9807565];
%
T.~Hahn,
Nucl.\ Phys.\ Proc.\ Suppl.\  {\bf 89} (2000) 231
[hep-ph/0005029].

\bibitem{Denner:2005es}
  A.~Denner, S.~Dittmaier, M.~Roth and L.H.~Wieders,
  Phys.\ Lett.\ B {\bf 612} (2005) 223
  [hep-ph/0502063].

\bibitem{'tHooft:1979xw}
G.~'t Hooft and M.~Veltman,
Nucl.\ Phys.\ B {\bf 153} (1979) 365;
%
W.~Beenakker and A.~Denner,
Nucl.\ Phys.\ B {\bf 338} (1990) 349;
%
A.~Denner, U.~Nierste and R.~Scharf,
Nucl.\ Phys.\ B {\bf 367} (1991) 637.

\bibitem{Denner:2002ii}
A.~Denner and S.~Dittmaier,
Nucl.\ Phys.\ B {\bf 658} (2003) 175
[hep-ph/0212259].

\bibitem{Passarino:1979jh}
G.~Passarino and M.~Veltman,
Nucl.\ Phys.\ B {\bf 160} (1979) 151.

\bibitem{Denner:2005nn}
  A.~Denner and S.~Dittmaier,
  Nucl.\ Phys.\ B {\bf 734} (2006) 62
  [hep-ph/0509141].

\bibitem{Ghinculov:1995bz}
  A.~Ghinculov,
  Nucl.\ Phys.\ B {\bf 455} (1995) 21
  [hep-ph/9507240];
%
  A.~Frink, B.A.~Kniehl, D.~Kreimer and K.~Riesselmann,
  Phys.\ Rev.\ D {\bf 54} (1996) 4548
  [hep-ph/9606310].

\bibitem{Dittmaier:2000mb}
S.~Dittmaier,
Nucl.\ Phys.\ B {\bf 565} (2000) 69
[hep-ph/9904440].

\bibitem{Bredenstein:2005zk}
  A.~Bredenstein, S.~Dittmaier and M.~Roth,
  Eur.\ Phys.\ J.\ C {\bf 44} (2005) 27
  [hep-ph/0506005].

\bibitem{Denner:2002cg}
  A.~Denner, S.~Dittmaier, M.~Roth and D.~Wackeroth,
  Comput.\ Phys.\ Commun.\  {\bf 153} (2003) 462
  [hep-ph/0209330].
 
\bibitem{Bredenstein:2004ef}
  A.~Bredenstein, S.~Dittmaier and M.~Roth,
  Eur.\ Phys.\ J.\ C {\bf 36} (2004) 341
  [hep-ph/0405169].

\bibitem{Lepage:1977sw}
  G.P.~Lepage,
  J.\ Comput.\ Phys.\  {\bf 27} (1978) 192.

\bibitem{Boos:2001cv}
  E.~Boos {\it et al.},
  arXiv:hep-ph/0109068.

\end{thebibliography}
\end{document}